\documentclass[pra,aps,twocolumn,showpacs]{revtex4}
\usepackage{epsfig} 
\begin{document}
\title{Precession and interference in the Aharonov-Casher 
and scalar Aharonov-Bohm effect}
\author{Philipp Hyllus$^{1}$\footnote{Electronic address: 
hyllus@itp.uni-hannover.de} and 
Erik Sj\"{o}qvist$^{2,}$\footnote{Electronic address: 
eriks@kvac.uu.se}} 
\affiliation{$^{1}$ Institut f\"{u}r Theoretische Physik, 
Universit\"{a}t Hannover, 30167 Hannover, Germany}  
\affiliation{$^{2}$Department of Quantum Chemistry, Uppsala 
University, Box 518, Se-751 20 Uppsala, Sweden} 
\date{\today}
\begin{abstract}
The ideal scalar Aharonov-Bohm (SAB) and Aharonov-Casher 
(AC) effect involve a magnetic dipole pointing in a certain 
fixed direction: along a purely time dependent magnetic field 
in the SAB case and perpendicular to a planar static electric 
field in the AC case. We extend these effects to arbitrary 
direction of the magnetic dipole. The precise conditions for 
having nondispersive precession and interference effects in 
these generalized set ups are delineated both classically 
and quantally. Under these conditions the dipole is affected 
by a nonvanishing torque that causes pure precession around 
the directions defined by the ideal set ups. It is shown 
that the precession angles are in the quantal case linearly related 
to the ideal phase differences, and that the nonideal phase 
differences are nonlinearly related to the ideal phase 
differences. It is argued that the latter nonlinearity is 
due the appearance of a geometric phase associated with 
the nontrivial spin path. It is further demonstrated that 
the spatial force vanishes in all cases except in the 
classical treatment of the nonideal AC set up, where the 
occurring force has to be compensated by the experimental 
arrangement. Finally, for a closed space-time loop the local 
precession effects can be inferred from the interference 
pattern characterized by the nonideal phase differences 
and the visibilities. It is argued that this makes it 
natural to regard SAB and AC as essentially local and 
nontopological effects. 
\end{abstract}
\maketitle
\section{Introduction}
\subsection{Key discoveries}
In the electric and magnetic Aharonov-Bohm (AB) \cite{aharonov59} 
effect a charged particle exhibits a physical 
effect due to an electromagnetic potential although the 
electromagnetic field and force vanish along the path of 
the particle. To each of these, a dual effect exists, the 
so-called scalar Aharonov-Bohm (SAB) effect and Aharonov-Casher 
(AC) effect, respectively. In the dual set ups, an electrically 
neutral particle carrying a nonvanishing magnetic dipole moment 
exhibits a physical effect due to an electromagnetic field 
although no force is acting on the particle. Despite this 
similarity, the physics behind SAB and AC differs crucially 
from that of the electric and magnetic AB effect. To discuss 
these differences and to enlighten the extra richness of the 
dual effects is the major aim of this paper.

\begin{figure}[ht!]
\begin{center}
\includegraphics[width=8 cm]{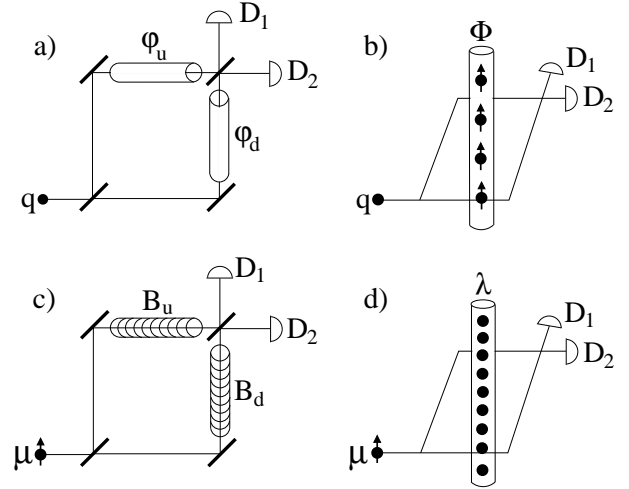}
\end{center}
\caption{The a) electric, b) magnetic, c) scalar Aharonov-Bohm 
and d) the Aharonov-Casher effect. Here, $q$ is the charge of the
particle in the AB effects, $\mbox{\boldmath $\mu$}$ is the magnetic
moment vector in the dual effects, and $D_{1,2}$ denote the detectors in
all set ups.  Furthermore, in a) $\varphi_{u,d}$ denote the electric
potential in the upper and the lower arm of the interferometer,
respectively, and $B_{u,d}$ in c) denote the magnetic fields in the
same manner.  In b) $\Phi$ is the magnetic flux caused by the solenoid
pictured by magnetic dipoles and $\lambda$ in d) is the line charge
density.  In the dual effects there is the restriction that
$\mbox{\boldmath $\mu$}\| {\bf B}$ in c) and $\mbox{\boldmath $\mu$}\|
{\bf e}_{\lambda}$ in d).}
\end{figure}

In the electric Aharonov-Bohm (EAB) effect (see Fig. 1a), 
a wave packet 
carrying electric charge $q$ is split coherently by a beam 
splitter, each beam passing through a cylindrical metal tube 
at vanishing electric potential. When the wave packets are 
well inside the tubes, the beam pair is exposed to a potential 
difference that varies as a pure function of time. Well before 
the wave packets leave the tubes, the potential vanishes again. 
The beams are brought together to interfere coherently 
at a second beam splitter. This process is essentially 
captured by the Hamiltonian 
\begin{equation}
H = q\varphi(t) , 
\label{eq:IntroEABham}
\end{equation}
where $\varphi$ is a function of time inside each tube. The 
interference of the wave packets depends on the phase difference 
between the two beams
\begin{equation}
\phi_{\text{EAB}} = -\frac{q}{\hbar} 
\int_{0}^{\tau} \Delta \varphi (t) \ dt,
\label{eq:IntroEABps}
\end{equation}
where $\Delta\varphi$ is the potential difference between
the metal tubes and $\tau$ is the time interval during 
which the electric potential is nonvanishing. Thus, there 
may be a physical effect on the particle although the 
electric field vanishes and no force is induced on its path 
by the time dependent potential. The effect is nondispersive, 
i.e. independent of the particle's velocity, as long as 
$\Delta \varphi (t)$ vanishes before the tail of the wave 
packets reaches the edges of the tubes.

In the magnetic Aharonov-Bohm (MAB) effect 
(see Fig. 1b), there is a long 
thin shielded solenoid perpendicular to the plane of motion 
and with a current flowing through it so as to produce a 
magnetic flux $\Phi$. A beam of particles with charge $q$ 
is split coherently into two beams that pass the solenoid 
on opposite sides and are brought together again to interfere. 
The solenoid has to be very long to make sure that the magnetic 
field ${\bf B}$ vanishes outside it. The vector potential 
${\bf A}$, however, cannot vanish everywhere outside the 
solenoid, because the line integral of ${\bf A}$ along any 
closed circuit $C$ that contains the solenoid equals the 
magnetic flux $\Phi$. The Hamiltonian relevant for the MAB 
case is
\begin{equation}
H = \frac{1}{2m} ({\bf p}-q{\bf A})^{2},
\end{equation}
where $m$ is the mass of the particle. In this set up the 
phase difference is given by
\begin{equation}
\phi_{\text{MAB}}
= \frac{q}{\hbar}\oint_{C} {\bf A}\cdot d{\bf r}
= \frac{q\Phi}{\hbar},
\end{equation}
so there may be a physical effect on the particle although 
the magnetic field vanishes and thus no force is induced 
along its path. Furthermore, the MAB phase difference is 
nondispersive as it is independent of the particle's velocity. 

To each of the described above, a dual effect exists: the scalar 
Aharonov-Bohm (SAB) effect for EAB and the Aharonov-Casher 
(AC) effect for MAB. Instead of charged particles, the dual 
effects involve electrically neutral particles carrying a 
magnetic moment $\mu$, e.g. neutrons, and the gauge potentials 
are replaced by magnetic and electric fields. By choosing certain 
configurations a complete mathematical analogy can be achieved 
in both cases. 

The ideal SAB set up (see Fig. 1c), 
first suggested by Zeilinger \cite{zeilinger86}, 
is similar to that of EAB. Instead 
of the electric potential, there is a spatially uniform 
magnetic field $B(t)$ in the $+z$ direction generated by 
a solenoid in each arm of the interferometer. The electrically 
neutral magnetic dipole points in the $+z$ direction and 
is split coherently. When the resulting wave packets are 
well inside the solenoids, the beam pair is exposed to a 
magnetic field difference that varies as a pure function 
of time. Well before the wave packets leave the solenoids, 
the magnetic field falls back to vanish. This process is 
described by the Hamiltonian
\begin{equation}
H = -\mu B(t),
\label{eq:IntroSABham}
\end{equation}
where $B(t)$ is a pure function of time inside each 
solenoid and $\mu$ is the magnetic dipole moment. The 
phase difference between the upper and the lower beam 
is given by
\begin{equation}
\phi_{\text{SAB}} = \frac{\mu}{\hbar} 
\int_{0}^{\tau} \Delta B(t) dt,
\label{eq:IntroSABps}
\end{equation}
where $\Delta B(t)$ is the magnetic field difference 
between the two solenoids and $\tau$ is the time interval 
during which the magnetic fields are nonvanishing. Again 
there may be a physical effect although no force is exerted 
on the particle. The effect is nondispersive as long as the 
applied particle velocity is such that $\Delta B (t)$ vanishes 
before the tail of the wave packets reaches the edges of the 
solenoids.

Consider again the magnetic AB effect. A solenoid in 
the $+z$ direction can be viewed as a line of magnetic 
dipoles pointing in the $+z$ direction. In the AC effect 
\cite{aharonov84}, the role of the magnetic dipoles and 
the charged particle is interchanged 
(see Fig. 1d). The set up is 
the same as that of the magnetic AB effect, but the 
solenoid is replaced by a line of charge in the $+z$ 
direction and the charged particle is replaced by a 
particle that carries a magnetic dipole moment $\mu$. 
The charged line gives rise to a static $z$ independent 
electric field in the $x-y$ plane. The dipole points 
in the $+z$ direction. For this set up, we obtain the 
Hamiltonian in the $x-y$ plane as 
\begin{equation}
H = \frac{1}{2m} ( {\bf p}-\mu {\bf A} )^{2}_{x-y},
\end{equation}
where ${\bf A} = c^{-2}(-E_{y},E_{x})$ and ${\bf E} = 
(E_{x},E_{y})$ is the electric field of the line of 
charge and $c$ is the speed of light. 
The phase difference is given by
\begin{equation} 
\phi_{\text{AC}} = 
\frac{\mu}{\hbar} \oint_{C} {\bf A} \cdot d\/{\bf r} = 
\frac{\mu \lambda}{\hbar\epsilon_{0} c^{2}},
\label{eq:idealACphase}
\end{equation}
where $\lambda$ is charge per unit length and $C$ is the 
interferometer loop. Again, there may be a 
physical effect although no force is exerted on the 
particle, and the AC phase difference is nondispersive.

Both SAB and AC phase differences have been verified 
experimentally. Experiments regarding the phase 
difference for a fixed particle velocity of SAB have been carried 
out by Allman {\it et al.} \cite{allman92,allman93,allman99} 
and regarding AC by Cimmino {\it et al.} \cite{cimmino89}, 
both using neutrons. Badurek {\it et al.} \cite{badurek93} 
examined the nondispersivity of SAB, also with neutrons. 
Sangster {\it et al.} \cite{sangster93,sangster95} and 
G\"{o}rlitz {\it et al.} \cite{gorlitz95} examined the 
nondispersivity and dependence on the electric field of 
AC using atom interferometry.

Dowling {\it et al.} \cite{dowling99} 
elaborated two further effects connected to the magnetic AB and 
the AC effects, related to them by the Maxwell electromagnetic 
duality relations. The Maxwell dual of MAB is a magnetic 
monopole in the field of a line of electric dipoles and that 
of AC is an electric dipole in the field of a line of magnetic 
monopoles, where the latter is also known as the 
He-McKellar-Wilkens effect \cite{he93,wilkens94}.

\subsection{Nonlocality, topology, and nondispersiveness}

The deep physical differences of the AB effects and their 
duals can be illustrated by the debate about 
nonlocality and topology in this context. 
Zeilinger \cite{zeilinger86} 
observed that a main feature of the AB effects and their 
duals is that the phase differences are nondispersive, i.e. 
independent of the particle's velocity. Allman {\it et al.} 
\cite{allman92,allman93,allman99} defined a topological 
effect as one in which a nonvanishing phase difference 
occurs although no force is exerted on the particle. 
Peshkin \cite{peshkin99} showed that force free motion 
is a necessary condition for a nondispersive effect.

Peshkin and Lipkin \cite{peshkin95} pointed out that 
a definition of topological effects related only to 
nondispersivity and force free motion is problematic. 
To illustrate their point, they suggested that 
the magnetic field in SAB could be replaced by an optical 
phase shifter whose uniform index of refraction depends 
on time while the particle is inside some box in one 
arm of the interferometer. They claimed that the phase 
difference in such a set up could in principle be made 
independent of the particle's velocity over the 
experimental range. Thus, a force free effect due 
to an optical device, which has no deep physical 
significance, would then be topological regarding 
the criterion stated in \cite{allman92,allman93,allman99}.

They analysed the above described effects and came to 
the conclusion that both the electric and the magnetic 
AB effect are nonlocal in the sense that there exist no 
electromagnetic fields along the path of the charged 
particle and no exchange of physical quantities takes 
place. They called the AB effects topological in that 
they require the charged particle to be confined to a 
multiply connected space-time region and in that there 
is no objective way to relate a phase shift to any 
particular place or arm of the interferometer. This 
is due to the fact that only the differences in phase 
shifts between different paths is gauge invariant, 
but no measurable phase shift can be assigned to 
any place because the integral of an electromagnetic 
potential along any open space-time path is gauge 
dependent. 

They remarked that in the dual effects there exist 
electromagnetic fields along the paths of the beams. 
They claimed that the particle interacts with these 
fields via angular momentum fluctuations although no 
force is exerted on it. They concluded that both the 
SAB and the AC effect are local because there are local 
interactions and nontopological because the phase shifts 
depend on the local fields along the paths and can thus 
be related to a particular place.

To conclude, using the criteria in \cite{peshkin95}, the 
AB effects are both nonlocal and topological while the 
dual effects are local and nontopological. All the effects 
have in common that they are force free and thus nondispersive. 
All effects require a closed path to measure the phase 
difference.

\subsection{Nonideal version of the dual effects}

The dual effects have one extra degree of freedom that 
makes them interesting: the direction of the magnetic 
dipole \cite{goldhaber89}. Nevertheless, a thorough 
analysis of the precession angles and phase differences 
occurring in the nonideal set ups, where this direction may 
be arbitrary, is still missing. 

Some work have been devoted to the precession of the magnetic 
moment in the nonideal set ups, but not from this perspective. 
Han and Koh \cite{han92} analysed the AC effect in the rest 
frame of the magnetic dipole and derived the AC phase from 
the precession of a magnetic dipole in a magnetic field, but 
they did not treat the nonideal AC case. Peshkin and Lipkin 
\cite{peshkin95} used a similar point of view to conclude 
that the dual effects are local and nontopological as mentioned 
above. They also analyzed SAB in terms of a spin autocorrelation 
effect related to the dipole precession, and one of the present 
Authors \cite{sjoqvist00} has recently discussed an analogous 
correlation effect between the spin and spatial degrees of 
freedom that arises in the nonideal AC set up.

To discuss and enlighten the extra richness related to the 
direction of the magnetic dipole in SAB and AC is the major 
aim of this paper. In particular, we shall examine in detail 
the conditions under which nondispersive dynamics occur 
in the generalized set ups. Analyses of both a classical 
magnetic dipole and a quantal spin$-\frac{1}{2}$ particle 
with magnetic moment are carried out.  

\section{Classical magnetic dipole}
\label{sec:Classical}

\subsection{Hamiltonian}

Aharonov and Casher \cite{aharonov84} analyzed a situation 
where an electrically neutral particle exhibits a magnetic 
AB type effect in the presence of a fixed distribution of 
charge. In this section we extend their analysis to moving 
point charges, in order to have a common physical basis for 
the classical treatment of the AC and SAB effect. We start 
from the AC Lagrangian describing a system of charged particles, 
all with charge $q$, interacting with magnetic dipoles, 
all with magnetic dipole moment $\mu$, (leaving out purely 
electric-electric terms)
\begin{eqnarray}
L & = & \frac{1}{2}\sum_{k} \tilde{m}_{k}\tilde{v}_{k}^{2} + 
\frac{1}{2}\sum_{l} m_{l}v_{l}^{2} 
\nonumber \\ 
 & & + \sum_{k,l} q {\bf A} (\tilde{{\bf r}}_{k}-{\bf r}_{l})
\cdot (\tilde{{\bf v}}_{k}-{\bf v}_{l}) .  
\label{eq:Lagrangian2}
\end{eqnarray}
Here $\tilde{m}_{k}, \tilde{{\bf r}}_{k}, \tilde{{\bf v}}_{k}$ 
denote the mass, position, and velocity, respectively, of the 
charged particle $k$, and $m_{l},{\bf r}_{l},{\bf v}_{l}$ 
denote the corresponding quantities for dipole $l$. Because 
of the dependence of ${\bf A}$ on $(\tilde{{\bf r}}_{k} - 
{\bf r}_{l})$, the system possesses a symmetry when the 
role of $\tilde{{\bf r}}_{k}$ and ${\bf r}_{l}$ is reversed.

The vector potential at $\tilde{{\bf r}}$ due to a classical 
magnetic dipole \mbox{\boldmath $\mu$} situated at ${\bf r}$ 
may be written as (see, e.g., \cite{jackson75}, p. 182)
\begin{equation}
{\bf A}(\tilde{{\bf r}}-{\bf r}) = \frac{1}{4\pi\epsilon_{0}c^{2}}
\frac{\mbox{\boldmath $\mu$}\times (\tilde{{\bf r}}-{\bf r})}
{|\tilde{{\bf r}}-{\bf r}|^{3}}.
\label{eq:mdvecpot}
\end{equation}
The electric field at ${\bf r}$ due to a particle situated 
at $\tilde{{\bf r}}_{k}$ carrying the charge $q$ is
\begin{equation} 
{\bf E}_{k} \equiv {\bf E} ({\bf r} - \tilde{{\bf r}}_{k}) = 
\frac{q}{4\pi\epsilon_{0}} \frac{({\bf r} - \tilde{{\bf r}}_{k})}
{|{\bf r} - \tilde{{\bf r}}_{k}|^{3}}.
\end{equation}
Thus, the Lagrangian of a single magnetic dipole moving in the 
field of moving charges may be written as 
\begin{eqnarray}
L & = & \frac{1}{2}mv^{2} +\sum_{k} q 
\frac{1}{4\pi\epsilon_{0}c^{2}} \frac{\mbox{\boldmath $\mu$}\times 
(\tilde{{\bf r}}_{k}-{\bf r})}{|\tilde{{\bf r}}_{k}-{\bf r}|^{3}} 
\cdot(\tilde{{\bf v}}_{k}-{\bf v}) 
\nonumber\\
 & = & \frac{1}{2}mv^{2} + \mbox{\boldmath $\mu$}\cdot
\sum_{k}\frac{\tilde{{\bf v }}_{k}}{c^{2}}\times {\bf E}_{k} + 
\Big( \frac{\mbox{\boldmath $\mu$}}{c^{2}}\times\sum_{k} 
{\bf E}_{k} \Big) \cdot {\bf v} . 
\nonumber \\ 
\label{eq:Lagrangian3}
\end{eqnarray}
This may be further simplified by noting that the dipole 
is exposed to a magnetic field (see, e.g., \cite{jackson75}, 
p. 542)
\begin{equation}
{\bf B}({\bf r}) = 
\sum_{k} \frac{\tilde{{\bf v}}_{k}}{c^{2}} \times 
{\bf E}_{k}  
\end{equation}
due to the moving point charges. Thus, the Lagrangian 
(\ref{eq:Lagrangian3}) finally becomes 
\begin{equation}
L = \frac{1}{2}mv^{2} + 
\mbox{\boldmath $\mu$}\cdot{\bf B} + \Big(
\frac{\mbox{\boldmath $\mu$}}{c^{2}}\times{\bf E}
\Big)\cdot{\bf v} , 
\label{eq:Lagrangian4}
\end{equation}
where we have introduced the total electric field ${\bf E} = 
\sum_{k}{\bf E}_{k}$. 

The corresponding Hamiltonian is obtained by Legendre 
transforming the variables of the dipole, but treating 
the velocities of the charged particles as parameters. 
This yields 
\begin{equation}
H = \frac{1}{2m} \Big( {\bf p}-\frac{\mbox{\boldmath $\mu$}}{c^{2}} 
\times{\bf E} \Big)^{2} - \mbox{\boldmath $\mu$}\cdot{\bf B} , 
\label{eq:ACclHam}
\end{equation}
where 
\begin{equation} 
{\bf p} = \mbox{\boldmath $\nabla$}_{\bf v} L = 
m{\bf v} + \frac{\mbox{\boldmath $\mu$}}{c^{2}} 
\times{\bf E} . 
\end{equation}

Note that the only visible classical SAB and AC effects will 
be the precession of the magnetic dipole, for which any 
spatial Lorentz force is irrelevant. Therefore the particle
must not necessarily be electrically neutral. The path of 
such a charged dipole can in principle be controlled by the 
experimental set up, compensating the occurring forces.

\subsection{Scalar Aharonov-Bohm}

From Eq. (\ref{eq:ACclHam}) we see that the Hamiltonian of 
an electrically neutral magnetic dipole $\mbox{\boldmath $\mu$}$ 
in a pure magnetic field ${\bf B}$ is given by
\begin{equation}
H = \frac{{\bf p}^{2}}{2m} - \mbox{\boldmath $\mu$} 
\cdot {\bf B} . 
\end{equation}
In the nonideal SAB set up, there is a spatially uniform 
magnetic field ${\bf B}=B(t){\bf \hat{z}}$ and the magnetic 
dipole can point in any direction. The condition for force
free motion in the ideal case is that the magnetic field
is nonvanishing only when the particle is well inside the
solenoid. This leads to force free motion also in the nonideal
case: $\mbox{\boldmath $\nabla$}_{{\bf x}} 
\Big(\mu_{z}B(t)\Big) = 0$. However, there can be force pairs 
generating a torque on the magnetic dipole. It can be assumed 
that $\mbox{\boldmath $\mu$}=\Gamma {\bf S}$ \cite{goldstein80}, 
where $\Gamma=\bar{q}/2\bar{m}$. Here $\bar{q}$ is the charge and 
$\bar{m}$ the mass of the particles of the current that 
is associated with the 
magnetic dipole moment $\mu$ and the intrinsic angular momentum 
${\bf S}$. The Poisson bracket relations $\{S_{k},S_{l}\} = 
\epsilon_{klm}S_{m}$ yield
\begin{equation}
\dot{\mbox{\boldmath $\mu$}} = 
\Gamma \mbox{\boldmath $\mu$} \times {\bf B},
\end{equation}
which reduces to
\begin{equation}
\dot{\mbox{\boldmath $\mu$}} = 
\Gamma B(t)(\mu_{y},-\mu_{x},0)
\end{equation}
in the SAB set up.
With the initial conditions $\mbox{\boldmath $\mu$}(0) = 
\mu (\sin \theta ,0, \cos \theta)$ we obtain
\begin{eqnarray}
\mbox{\boldmath $\mu$} & = & 
\mu (\sin \theta \cos \gamma(t) , \sin \theta \sin \gamma(t) , 
\cos \theta ),
\nonumber \\ 
\gamma(t) & = & - \Gamma\int_{0}^{t} B(t')dt'.
\label{eq:SABcp}
\end{eqnarray}
The main point here is that the precession angle $\gamma(t)$ 
is independent of the particle's velocity as the particle 
does not feel a field gradient under the SAB conditions. 
This also assures that the angle $\theta$ with respect to 
the $z$ axis remains constant. 

\subsection{Aharonov-Casher}

From Eq. (\ref{eq:ACclHam}) we see that the Hamiltonian of a
magnetic dipole $\mbox{\boldmath $\mu$}$ in a pure electric 
field ${\bf E}$ is given by
\begin{equation}
H = \frac{1}{2m} \Big( {\bf p} - 
\frac{\mbox{\boldmath $\mu$}}{c^{2}} \times{\bf E} \Big)^{2},
\end{equation}
where $m$ is the mass of the dipole. Assuming again
$\mbox{\boldmath $\mu$}=\Gamma {\bf S}$, the precession 
and the net force acting on the magnetic dipole can be 
derived using Hamilton's equations
\begin{eqnarray}
\dot{\mbox{\boldmath $\mu$}} & = & 
- \frac{\Gamma}{c^{2}}\mbox{\boldmath $\mu$}
\times \Big({\bf v}\times {\bf E}\Big)
\label{eq:gentorque} \\
m{\bf \dot{v}} & = & {\bf v}\times
\Big( \mbox{\boldmath $\nabla$} \times \big( 
\frac{\mbox{\boldmath $\mu$}}{c^{2}}\times{\bf E} \big) 
\Big) = -\frac{{\bf v}\times (\mbox{\boldmath $\mu$} \cdot 
\mbox{\boldmath $\nabla$}){\bf E}}{c^{2}},
\label{eq:genforce}
\end{eqnarray}
where ${\bf v}$ denotes the velocity of the particle. 
The dipole sees in its own rest frame the magnetic field 
$-(1/c^{2})({\bf v}\times {\bf E})$, which causes the 
precession described by Eq. (\ref{eq:gentorque}). In 
Eq. (\ref{eq:genforce}) we used that 
$\mbox{\boldmath $\nabla$}\cdot{\bf E}$ vanishes 
where the particle moves. The conditions for the 
ideal AC effect to be force free and nondispersive
are that the particle moves in the $x-y$ plane and
that there is an electric field of the form 
\begin{equation}
{\bf E}=E_{x}(x,y){\bf \hat{x}}+E_{y}(x,y){\bf \hat{y}},
\label{eq:efield}
\end{equation}
which can be generated by a charge distribution $\rho (x,y)$ 
that is independent of $z$ and that is shielded from the 
dipole's path. In the nonideal case, we allow for arbitrary 
direction of the magnetic dipole. With these conditions 
we obtain a nonvanishing force only in the $z$ direction. 
In the classical case, $v_{z}=0$ is a constraint that has 
to be fulfilled for a nondispersive effect, thus making 
it necessary to compensate this force by the experimental 
set up. Under this constraint, the expression for the 
precession reduces to
\begin{equation}
\dot{\mbox{\boldmath $\mu$}} = 
\Gamma {\bf A} \cdot d{\bf v} (\mu_{y},-\mu_{x},0) , 
\end{equation}
where we defined ${\bf A}=c^{-2}(-E_{y},E_{x},0)$. 
These equations are solved by
\begin{eqnarray}
 \mbox{\boldmath $\mu$} & = & \mu (\sin \theta \cos 
\gamma({\bf x}) , \sin \theta \sin \gamma({\bf x}) , 
\cos \theta) ,
\nonumber \\ 
\gamma({\bf x}) & = & 
-\Gamma \int_{{\bf x}_{0}}^{\bf x} {\bf A} \cdot d{\bf r}  
\label{eq:ACcp}
\end{eqnarray}
with ${\mbox{\boldmath $\mu$}} ({\bf x}_{0}) = 
\mu (\sin \theta , 0 , \cos \theta )$ at the reference point 
${\bf x}_{0}$ in the $x-y$ plane. Thus, the precession is 
velocity independent, while the angle $\theta$ with respect 
to the $z$ axis remains constant. The precession is independent 
of the details of the path since $\mbox{\boldmath $\nabla$} 
\times {\bf A} \propto \mbox{\boldmath $\nabla$}\cdot{\bf E}$ 
vanishes where the particle moves. For a closed loop Stokes' 
theorem yields the precession angle
\begin{equation}
\gamma_{\text{AC}} = 
-\Gamma \oint_{C = \delta S} {\bf A} \cdot d{\bf r} = 
-\frac{\Gamma}{c^{2}} \int_{S} (\mbox{\boldmath $\nabla$} 
\cdot {\bf E}) dS = -\frac{\Gamma\lambda}{\epsilon_{0}c^{2}},
\end{equation}
independent of the velocity or the details of the
spatial path of the particle. Here
$\lambda$ denotes the linear charge density enclosed 
by the path $C$. 

\section{Spin$-\frac{1}{2}$}
\label{sec:Quantal}

\subsection{Hamiltonian}

Consider an electrically neutral spin$-\frac{1}{2}$ particle 
of rest mass $m$ carrying a magnetic moment $\mu$ and moving 
in an electromagnetic field described by the field tensor 
$F^{\kappa\delta} = \partial^{\kappa} A^{\delta} - 
\partial^{\delta} A^{\kappa}$, $A^{\kappa}$ being the 
corresponding four-potential. This system obeys the Dirac 
equation \cite{pauli41,foldy52}
\begin{equation}
\left( i\hbar \gamma^{\rho}\partial_{\rho} - 
mc - \frac{\mu}{2c^{2}} F_{\kappa\delta} 
\sigma^{\kappa\delta}\right)\psi = 0 , 
\label{eq:relativity}
\end{equation}
which takes the form 
\begin{equation}
i\hbar\frac{\partial}{\partial t}\psi = 
\left( c \mbox{\boldmath $\alpha$} \cdot {\bf p} + 
\beta mc^{2} -\beta\mu\mbox{\boldmath $\sigma$}\cdot{\bf B} + 
i \frac{\mu}{c} \mbox{\boldmath $\gamma$} \cdot {\bf E} \right) \psi 
\label{eq:dirac2}
\end{equation}
by rewriting the Dirac matrices as $(\gamma^{0} , \gamma^{k}) = 
(\beta ,\beta\alpha_{k})$ and choosing the representation 
\cite{bjorken64}
\begin{equation}
\alpha_{k} = \left[ \begin{array}{cc}
0 & \sigma_{k} \\
\sigma_{k}&0 
\end{array} \right],
\hspace{1cm}
\beta = \left[ \begin{array}{cc}
I & 0 \\
0 & -I 
\end{array} \right] . 
\end{equation}
Here, $\sigma_{k}$ are the standard Pauli matrices defining 
$\mbox{\boldmath $\sigma$} = (\sigma_{x},\sigma_{y},\sigma_{z})$, 
$I$ is the $2 \times 2$ unit matrix, ${\bf E}$ and ${\bf B}$ 
denote the electric and the magnetic field, respectively, 
and we have used that $\sigma^{\mu\nu}= \frac{i}{2} 
[\gamma^{\mu},\gamma^{\nu}]$. 

In order to take the nonrelativistic limit it is convenient to 
express the four-spinor $\psi$ in terms of the two-spinors 
$\varphi$ and $\chi$
\begin{equation} 
\psi = e^{-i(mc^{2}/\hbar)t} \Big( \begin{array}{c} 
\varphi \\ 
\chi 
\end{array} \Big)
\end{equation}
yielding the coupled equations 
\begin{eqnarray}
i \hbar \frac{\partial}{\partial t} 
\Big( \begin{array}{c} 
\varphi \\ \chi 
\end{array} \Big)
 & = & c \mbox{\boldmath $\sigma\cdot p$}
\Big( \begin{array}{c} 
\chi \\ \varphi 
\end{array} \Big)
-2mc^{2} 
\Big( \begin{array}{c} 
0 \\ \chi 
\end{array}\Big)
\nonumber \\ 
 & & -\mbox{\boldmath $\mu$}\cdot {\bf B}
\Big( \begin{array}{r} 
\varphi \\ -\chi 
\end{array} \Big) 
+ i \frac{1}{c} \mbox{\boldmath $\mu$} \cdot {\bf E}
\Big( \begin{array}{r} \chi \\ -\varphi 
\end{array} \Big) , 
\label{eq:dirac4}
\end{eqnarray}
where we have defined $\mbox{\boldmath $\mu$} = 
\mu\mbox{\boldmath $\sigma$}$. The lower of these 
equations is
\begin{equation}
\Big( i \hbar\frac{\partial}{\partial t} - 
\mbox{\boldmath $\mu$}\cdot {\bf B} + 2mc^{2} \Big) \ \chi = 
c\mbox{\boldmath $\sigma$} \cdot \Big( {\bf p} - 
i \frac{\mu}{c^{2}} {\bf E} \Big) \ \varphi.
\label{eq:lowerequation}
\end{equation}
In the nonrelativistic regime, the terms $i\hbar 
(\partial \chi /\partial t)$ and $\mbox{\boldmath $\mu$} 
\cdot {\bf B}\ \chi$ are negligible compared to the rest energy 
term $mc^{2}\ \chi$. Thus, we obtain the approximate expression 
\begin{equation}
\chi = \frac{1}{2mc} \mbox{\boldmath $\sigma$} \cdot 
({\bf p} -i \frac{\mu}{c^{2}} {\bf E}) \ \varphi .
\label{eq:smallcomponents}
\end{equation}
Note that $\chi$ is reduced by $\sim v/c\ll 1$ compared to $\varphi$ in the 
nonrelativistic regime and can therefore be neglected. We 
insert Eq. (\ref{eq:smallcomponents}) into the upper of
Eqs. (\ref{eq:dirac4}) and obtain the two-spinor equation
\begin{eqnarray}
i \hbar \frac{\partial}{\partial t} \ \varphi & = & 
\frac{1}{2m} \left(
\mbox{\boldmath $\sigma$}\cdot ({\bf p}+i\frac{\mu}{c^{2}}{\bf E})\ 
\mbox{\boldmath $\sigma$}\cdot ({\bf p}-i\frac{\mu}{c^{2}}{\bf E})
\right) \varphi
\nonumber \\ 
 & & -\mbox{\boldmath $\mu$} \cdot {\bf B} \ \varphi.
\label{eq:dirac5}
\end{eqnarray} 
Using the identity $(\mbox{\boldmath $\sigma$} \cdot 
{\bf a}) \ (\mbox{\boldmath $\sigma$} \cdot {\bf b}) = 
{\bf a} \cdot {\bf b} + i \mbox{\boldmath $\sigma$} 
\cdot ({\bf a} \times {\bf b})$ we finally obtain 
\cite{anandan89,he91}
\begin{eqnarray}
i \hbar \frac{\partial}{\partial t} \ \varphi & = & 
\Big[ \frac{1}{2m} \left( 
{\bf p} - \frac{\mbox{\boldmath $\mu$}}{c^{2}} \times {\bf E}
\right)^{2}
-\frac{\mu\hbar}{2mc^{2}} \mbox{\boldmath $\nabla$} \cdot {\bf E}
\nonumber \\ 
 & & -\frac{\mu^{2}E^{2}}{2mc^{4}} - 
\mbox{\boldmath $\mu$}\cdot {\bf B} \Big] \varphi ,
\label{eq:nonrellimit}
\end{eqnarray}
which is the Schr\"{o}dinger equation responsible for the 
SAB and AC effect. In the following, we use the standard 
nonrelativistic notation and put $\varphi \equiv |\Psi\rangle$. 

\subsection{Scalar Aharonov-Bohm}

In the SAB set up, Eq. (\ref{eq:nonrellimit}) reduces to
\begin{equation}
i\hbar \frac{\partial}{\partial t} |\Psi\rangle = 
\Big[ \hat{I} \otimes \frac{\hat{\bf p}^{2}}{2m} - 
\mu \hat{\sigma}_{z} \otimes B(\hat{\bf x},t) \Big]
|\Psi\rangle .
\end{equation}
Here and in the following, tensorial product distinguishes 
the spatial and the spin degrees of freedom, and hats 
distinguish operators from their expectation values. 
The conditions for the ideal SAB effect to be force free are 
that the magnetic field is nonvanishing only in a certain 
spatial region and during a certain time interval $\tau$, 
when the particle is well inside that region. The time 
development of $|\Psi\rangle$ is given by 
\begin{equation}
|\Psi(t)\rangle = 
\exp \Big[ -\frac{i}{\hbar} \hat{I} \otimes \hat{H}_{0} t + 
\frac{i}{\hbar} \hat{\sigma}_{z} \otimes \int_{0}^{t} 
B(\hat{\bf x},t') dt \Big] |\Psi_{0}\rangle,
\end{equation}
where $\hat{H}_{0}=\hat{\bf p}^{2}/2m$. In general, 
$\mbox{\boldmath $\nabla$}_{\hat{\bf x}} B(\hat{\bf x},t)$ 
does not vanish since there is a gradient at the edge points 
of the magnetic field region. The SAB conditions ensure that
the particle does not feel that gradient. Thus, although 
the quantum force $\mu \hat{\sigma}_{z} \otimes 
\mbox{\boldmath $\nabla$}_{\hat{\bf x}} B (\hat{\bf x},t)$ 
does not vanish everywhere, its expectation value vanishes due 
to the SAB conditions. Therefore, the particle moves force 
free and we may effectively write $B(\hat{\bf x},t) = 
B(t) \hat{I}$. This holds for arbitrary direction of the 
magnetic dipole.

Initially, the state is represented by the product vector 
$|\Psi_{0}\rangle = |s_{0} \rangle \otimes | \psi_{0} \rangle$, 
where $|s_{0}\rangle$ is the spin vector and $|\psi_{0}\rangle$ 
is the spatial vector. At a later instant of time, the wave 
function in position representation reads
\begin{eqnarray}
\langle {\bf x}|\Psi(t)\rangle & = & 
\exp \Big[ -\frac{i}{2} \hat{\sigma}_{z} \gamma(t) \Big]
|s_{0}\rangle\ \psi ({\bf x},t) , 
\label{eq:SABwave}
\end{eqnarray}
where the spatial part $\psi ({\bf x},t) = \langle{\bf x}| 
\exp \big[ -(i/\hbar) \hat{H}_{0}t \big] |\psi_{0} \rangle$ is 
a scalar (spin independent) free particle wave packet that 
is independent of the magnetic field under the SAB conditions. 
The operator $\exp [-i\hat{\sigma}_{z}\gamma/2]$ rotates the 
spin around the $z$ axis with the angle
\begin{equation}
\gamma(t) = -2 \frac{\mu}{\hbar} \int_{0}^{t} B(t') dt'
\label{eq:SABqp}
\end{equation} 
when the particle moves in the magnetic field.

The precession angle $\gamma$ could be related to the ideal 
phase difference given by Eq. (\ref{eq:IntroSABps}) in the following 
way: First let the magnetic dipole pass the upper solenoid 
forward in time and take it back to its starting point 
through the lower solenoid backward in time, which yields 
the precession
\begin{eqnarray}
\gamma_{\text{SAB}}&=&-2\frac{\mu}{\hbar}
\Big(\int_{0}^{\tau}B_{u}(t)dt+\int_{\tau}^{0}B_{l}(t)dt\Big)
\nonumber\\
&=&-2\frac{\mu}{\hbar}\int_{0}^{\tau} \Delta B(t)dt
  =-2\phi_{\text{SAB}},
\end{eqnarray}
where $\tau$ is the time interval in which the magnetic
field is nonvanishing and the factor $2$ is the usual rotation 
factor for spin$-\frac{1}{2}$. The spin precession effect 
is nondispersive as the precession angle takes the value 
$-2\phi_{\text{SAB}}$ for all velocities, as long as the 
SAB conditions are fulfilled. In the ideal SAB case, where 
$|s_{0}\rangle$ is an eigenstate of $\hat{\sigma}_{z}$, there 
is no precession effect, but the interference pattern is 
shifted by $\phi_{\text{SAB}}$.

\subsection{Aharonov-Casher}
\label{sec:ACQ}
In the AC set up a neutral spin-$\frac{1}{2}$ particle with a 
magnetic dipole moment $\mu$ is moving in a plane, say the 
$x-y$ plane, where a nonvanishing external static electric 
field $\hat{\bf E} = {\bf E} (\hat{\bf x})$ that fulfills 
$\mbox{\boldmath $\nabla$}_{\hat{\bf x}} \cdot \hat{\bf E} = 0$ 
exists and where the magnetic field vanishes. Thus, Eq. 
(\ref{eq:nonrellimit}) reduces to
\begin{eqnarray}
i \hbar \frac{\partial}{\partial t} |\Psi\rangle 
 & = & \Big[ \frac{1}{2m} \left( 
\hat{I} \otimes\hat{\bf p} - \frac{\mu}{c^{2}} 
\hat{\mbox{\boldmath $\sigma$}} \otimes \times \hat{\bf E}
\right)^{2}
\nonumber \\ 
 & & -\hat{I} \otimes \frac{\mu^{2}\hat{E}^{2}}{2mc^{4}}
\Big] |\Psi\rangle , 
\label{eq:ACqHam1}
\end{eqnarray}
where again the tensorial product distinguishes the spin 
and spatial degrees of freedom. The ideal AC set up becomes force 
free and nondispersive if $\hat{E}_{z}=0, \ \partial_{\hat{z}} 
\hat{\bf E} = 0 \mbox{, and } \hat{p}_{z} |\Psi\rangle = 0$, 
where $|\Psi\rangle$ is the total state vector. Under these 
conditions, Eq. (\ref{eq:ACqHam1}) reduces to the 
Schr\"{o}dinger equation in the $x-y$ plane 
\cite{hagen90,he91}
\begin{equation}
i \hbar \frac{\partial}{\partial t} |\Psi \rangle = 
\frac{1}{2m} \Big( \hat{I} \otimes \hat{\bf p} - 
\mu \hat{\sigma}_{z} \otimes \hat{\bf A} \Big)^{2} 
|\Psi\rangle,
\label{eq:ACHam}
\end{equation}
where from now on $\hat{\bf p}=(\hat{p}_{x},\hat{p}_{y})$ 
and $\hat{\bf A} = c^{-2}(-\hat{E}_{y},\hat{E}_{x})$. Note 
that these conditions are independent of the spin state and 
the resulting Schr\"{o}dinger equation in the $x-y$ plane 
should also be valid in the nonideal case. As a result, the 
motion of the particle under this Hamiltonian is force free, 
contrary to the classical treatment of the nonideal AC set 
up. This can be seen by writing $|\Psi\rangle$ as  
\begin{equation}
|\Psi \rangle = \hat{\cal{U}} |\Psi' \rangle , 
\hspace{0.5cm}
\hat{\cal{U}} = e^{-\frac{i}{2} \hat{\sigma}_{z} \otimes 
\hat{\gamma}},
\hspace{0.5cm}
\hat{\gamma} = \gamma (\hat{{\bf x}}),
\label{eq:rotation}
\end{equation}
identifying $\hat{\cal{U}}$ as the spin rotation operator
around the $z$ axis with the local operator $\hat{\gamma}$ 
that rotates the spin state the angle $\gamma (\hat{{\bf x}})$.
Now, the 
vector $|\Psi'\rangle$ evolves according to the 
Schr\"{o}dinger-like equation
\begin{equation}
i \hbar \frac{\partial}{\partial t} |\Psi' \rangle = 
\hat{H}' |\Psi' \rangle ,
\label{eq:ACSchrod2} 
\end{equation} 
where the transformed Hamiltonian operator $\hat{H}' = 
\hat{\cal{U}}^{\dagger} \hat{H} \hat{\cal{U}}$ is given by 
\begin{equation}
\hat{H}' = 
\frac{1}{2m} \Big( \hat{I} \otimes \hat{\bf p}
-\hat{\sigma}_{z} \otimes \Big[ \frac{\hbar}{2} 
\mbox{\boldmath $\nabla$}\hat{\gamma} + 
\mu\hat{\bf A} \Big] \Big)^{2}.
\end{equation}
This reduces to the free particle Hamiltonian 
$\hat{H}_{0}=\hat{\bf p}^{2}/2m$ in the 
$x-y$ plane if we set 
\begin{equation}
\gamma(\hat{\bf x})=-\frac{2\mu}{\hbar}
\int^{\hat{\bf x}}{\bf A}\cdot d{\bf r} ,
\label{eq:ACqp}
\end{equation}
which can be done because $\partial_{x} A_{y} - \partial_{y} 
A_{x}=(1/{c^{2}}) \mbox{\boldmath $\nabla$}\cdot{\bf E}$ 
vanishes along the path of the particle. Let us further assume  
initially that $|\Psi_{0}'\rangle = |s_{0}\rangle \otimes 
|\psi_{0}\rangle$, where $|s_{0}\rangle$ is the spin and 
$|\psi_{0}\rangle$ is the spatial state vector. The subsequent 
AC wave function in position representation is given by
\begin{eqnarray}
\langle{\bf x}|\Psi(t)\rangle & = & 
\langle{\bf x}| \exp \Big[ -\frac{i}{2} \hat{\sigma}_{z} 
\otimes \gamma (\hat{\bf x}) \Big] |s_{0}\rangle 
\nonumber\\
 & & \otimes \exp \Big[ -\frac{i}{\hbar}\hat{H}'t \Big]
|\psi_{0}\rangle 
\nonumber\\
 & = & 
\exp \Big[ -\frac{i}{2}\hat{\sigma}_{z} \gamma ({\bf x}) \Big]  
|s_{0}\rangle \psi ({\bf x},t) 
\label{eq:ACwave}
\end{eqnarray}
where $\psi ({\bf x},t) = \langle{\bf x}| \exp \big[ 
-(i/\hbar) \hat{H}'t \big] |\psi_{0} \rangle$ is a scalar 
(spin independent) free particle wave packet and we have 
used that $\hat{\gamma}$ is a local operator so that 
$\langle {\bf x} | f[\hat{\gamma}] | {\bf x}' \rangle = 
f[ \gamma ({\bf x})] \delta ({\bf x}' - {\bf x})$ for any 
sufficiently well-behaved function $f$. Thus, the spin 
part is precessing the angle $\gamma({\bf x})$ independent 
of the path or the velocity of the particle. After the wave 
packet $\psi ({\bf x},t)$ has traversed a loop enclosing 
the line charge $\lambda$, Stokes' theorem yields 
the precession angle
\begin{equation}
\gamma_{\text{AC}}=
-\frac{2\mu}{\hbar}\oint_{C=\partial S} {\bf A}\cdot d{\bf r} 
= \frac{2\mu\lambda}{\hbar c^{2}\epsilon_{0}}
=-2\phi_{\text{AC}} , 
\end{equation}
where again the factor $2$ is the rotation factor for 
spin$-\frac{1}{2}$. In the ideal AC case, where 
$\langle s_{0} | \hat{\sigma}_{z} |s_{0} \rangle = +1$, 
there is no precession effect, but the interference pattern 
is shifted by $\phi_{\text{AC}}$.

In the AC set up, the region outside the charge is multiply 
connected. Under the condition that at any instant of time 
the wave function does not enclose this charge, the wave 
function (\ref{eq:ACwave}) is still single-valued. In 
interferometry \cite{cimmino89}, it is reasonable to assume 
that this condition is fulfilled if the particle's spatial 
wave function is a quantal wave packet, given the macroscopic 
size of the interferometer.

\subsection{Phase difference}

In the interferometer set up, both beams are influenced 
by magnetic fields for SAB and by an electric field in 
the $x-y$ plane for AC (see Fig. 2). 
The resulting phase difference 
between the state vector $|\Psi_{u}\rangle$ that went the 
upper path and $|\Psi_{d}\rangle$ that went the lower path 
for arbitrary polarization is given by the Pancharatnam 
connection \cite{pancharatnam56}
\begin{equation}
\phi = \arg \langle \Psi_{d} | \Psi_{u} \rangle.
\label{eq:Panch1}
\end{equation} 
This phase and the corresponding visibility $|\langle \Psi_{d} | 
\Psi_{u}\rangle|$ can be measured by adding an additional phase 
$\chi$ to one of the beams \cite{wagh95,wagh98}. 

\begin{figure}[ht!]
\begin{center}
\includegraphics[width=8 cm]{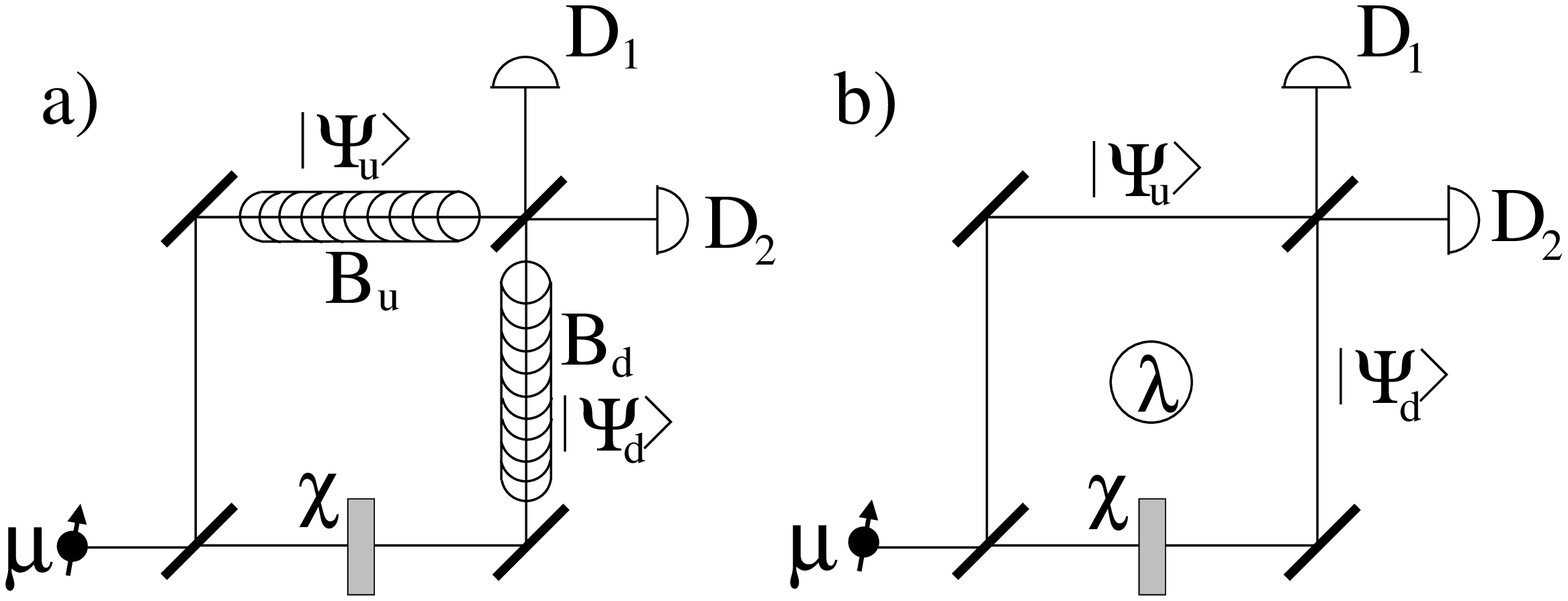}
\end{center}
\caption{Interferometer set ups for the nonideal versions
of a) the scalar Aharonov-Bohm and b) the Aharonov-Casher 
effect.  Here $\mbox{\boldmath $\mu$}$ is the magnetic 
moment vector, $D_{1,2}$ denote the detectors and $\chi$
is the extra $U(1)$ phase shift. Furthermore, 
$|\Psi_{u,d}\rangle$ and $B_{u,d}$ are the states 
and the magnetic fields in the upper and lower path, 
respectively. In b) $\lambda$ is the line charge density 
perpendicular to the interferometer plane.}
\end{figure}

Now, let us assume that the outgoing pair of spatial wave 
packets both have the same shape after leaving each 50-50 
beam splitter and that the free particle spreading is the 
same in both paths due to symmetry. Under these conditions 
it follows from Eqs. (\ref{eq:SABwave}) and (\ref{eq:ACwave}) 
that the wave function after the final beam splitter can be 
written as
\begin{eqnarray}
\langle {\bf x}|\Psi\rangle & = & 
\frac{1}{2} \Big( |s_{u} \rangle + e^{i\chi} 
|s_{d} \rangle \Big) \otimes\psi_{D_{1}} 
\nonumber \\ 
 & & + \frac{1}{2} \Big( |s_{u} \rangle - e^{i\chi} 
|s_{d} \rangle \Big) \otimes \psi_{D_{2}} ,
\end{eqnarray}
where $\psi_{D_{1,2}}$ correspond to the spatial wave 
functions at the detectors $1$ and $2$, respectively, and 
the phase $\chi$ was applied to the lower ($d$) beam. 
The spin parts are given by
\begin{equation}
|s_{u,d}\rangle = \exp \Big[ -\frac{i}{2} \hat{\sigma}_{z} 
\gamma_{u,d} \Big] |s_{0}\rangle,
\end{equation}
where the angle $\gamma$ is given by Eq. (\ref{eq:SABqp}) 
for SAB and by Eq. (\ref{eq:ACqp}) for AC. Thus, Eq. 
(\ref{eq:Panch1}) reduces to
\begin{equation}
\phi = \arg \langle s_{d} | s_{u} \rangle = 
\arg \langle s_{0} | \exp \Big[ \frac{i}{2} \hat{\sigma}_{z} 
(\gamma_{d} - \gamma_{u}) \Big] | s_{0} \rangle 
\end{equation} 
and we obtain the following detection probabilities 
for the detectors $D_{1}$ and $D_{2}$:
\begin{eqnarray}
P_{1} & = & 
\frac{1}{2} \Big[ 1 + \nu \cos (\phi + \chi) \Big] ,
\nonumber\\
P_{2} & = & 
\frac{1}{2} \Big[ 1 - \nu \cos (\phi + \chi) \Big] .
\label{eq:Detectors}
\end{eqnarray}

The visibility $\nu \equiv |\langle s_{d}|s_{u}\rangle|$ and 
the phase difference $\phi$ can be obtained experimentally by 
varying $\chi$. Without loss of generality, the initial spin 
state vector can be written as
\begin{equation}
\label{eq:GenSpin}
|s_{0}\rangle = \cos\frac{\theta}{2} |+\rangle + 
\sin\frac{\theta}{2}|-\rangle ,
\end{equation}
where the spin direction makes an angle $\theta$ with
the $z$ axis and its angle relative to the $x$ axis 
vanishes. For $\theta \neq \pi /2$, we obtain the phase 
difference and visibility 
\begin{eqnarray} 
\phi & = & \arctan{\Big(\cos{\theta}\tan{\phi_{D}} \Big)} , 
\label{eq:phasediff1} \\
\nu & = & \sqrt{1-\sin{^{2}\theta}\sin{^{2}\phi_{D}}} ,
\label{eq:visibility1}
\end{eqnarray}
being nonlinearly related to the ideal phase differences 
\begin{equation}
\phi_{D}=(\gamma_{d}-\gamma_{u})/2
=\frac{\mu}{\hbar} \int_{0}^{\tau} \Delta B dt = \phi_{{\text{SAB}}}
\end{equation}
in the SAB case and
\begin{equation}
\phi_{D} = \frac{\mu}{\hbar} 
\int_{u}^{\bf x}{\bf A}\cdot d{\bf r} - 
\frac{\mu}{\hbar} \int_{d}^{\bf x} 
{\bf A}\cdot d{\bf r} = \phi_{{\text{AC}}}
\end{equation} 
in the AC case. Similarly, for $\theta = \pi /2$ the phase 
shift and visibility become 
\begin{eqnarray} 
\phi & = & \left\{ \begin{array}{ll} 
0 & \ {\text{for}} \ 0 \leq \phi_{D} < \pi/2 \\
{\text{undefined}} & \ {\text{for}} \ \phi_{D} = \pi/2 \\ 
\pi & \ {\text{for}} \ \pi/2 < \phi_{D} \leq \pi 
\end{array} \right. , 
\label{eq:phasediff2} \\
\nu & = & |\cos \phi_{D} | .
\label{eq:visibility2}
\end{eqnarray}

Conversely, it should be noted that by observing the phase 
difference and visibility in the nonideal case, one may infer 
the spin precession effect. This can be seen by inverting 
Eqs. (\ref{eq:phasediff1}) and (\ref{eq:visibility1})
yielding 
\begin{eqnarray} 
\tan^{2} \phi_{D} = 
\frac{1}{\nu^{2} \cos^{2} \phi} - 1 
\nonumber \\ 
\cos^{2} \theta = 
\nu^{2} \frac{1-\cos^{2} \phi}{1-\nu^{2}\cos^{2} \phi},  
\end{eqnarray} 
which is also consistent with 
Eqs. (\ref{eq:phasediff2}) and (\ref{eq:visibility2}). 
Thus, measuring the interference pattern in fact determines 
the precession and thereby confirms the total accumulation 
of local torque around the interferometer loop.
This makes it natural to regard also the ideal phase 
differences $\phi_{{\text{SAB}} / {\text{AC}}}$ as 
essentially local quantities originated by the local 
torque and in turn the dual effects as nontopological. 
This is in contrast to the view of Aharonov and Reznik 
\cite{aharonov00}, who regard the phase shifts accumulated 
along the beam paths of SAB and AC as nonlocal quantities 
and suggest a complementarity between these phase shifts 
and the precession of the dipole moment, and led to further 
discussion in Refs. \cite{hyllus02,aharonov02}.

The phase differences in Eqs. (\ref{eq:phasediff1}) and 
(\ref{eq:phasediff2}) between the beam pair comprise a 
dynamical contribution 
\begin{equation} 
\gamma_{d} = \phi_{D} \cos \theta 
\end{equation}
that is directly proportional to the projection of 
the dipole moment onto the direction of the magnetic 
field in SAB and the axis of the enclosed line charge 
in AC. In addition, there is a geometrical contribution 
to the phase shift given by 
\begin{equation} 
\gamma_{g} = \phi - \phi_{D} \cos \theta = 
-\frac{1}{2} \Omega_{g-c} , 
\end{equation}
which is proportional to the geodesically closed solid 
angle $\Omega_{g-c}$ defined by the spin path and the 
shortest geodesic on the Bloch sphere. The phase difference 
between the upper and the lower beam is directly proportional 
to the ideal phase shifts under the condition that the 
geometric phase vanishes, which happens if $\theta$ is a 
multiple of $\pi$, i.e. the ideal case, where the spin 
state is an eigenstate of $\hat{\sigma}_{z}$ and no 
precession occurs. Thus, from this point of view the 
ideal phase shifts $\phi_{{\text{SAB}}/{\text{AC}}}$ 
are dynamical in nature and their nonlinear relation to 
the observed phase shift $\phi$ in the nonideal case may 
be understood from the occurrence of a geometric phase 
associated with the nontrivial line bundle $SU(2)/U(1)$ 
of spin$-\frac{1}{2}$. 

\section{Conclusions}
\label{sec:Conclusions}

We have extended the scalar Aharonov-Bohm (SAB) effect and 
the Aharonov-Casher (AC) effect to arbitrary directions of 
the magnetic dipole. The precise conditions for having 
nondispersive precession and interference effects in these 
generalized set ups have been delineated both classically 
and quantally. Under these conditions the dipole is affected 
by a nonvanishing torque that rotates its direction on a 
cone with an angle $\gamma$ around the directions defined 
by the ideal set ups. For a closed clockwise space-time 
interferometer loop, this angle is in the quantal case 
related to the ideal phase differences as $\gamma = 
-2\phi_{\text{SAB/AC}}$. The force vanishes in all cases 
except in the classical treatment of the nonideal AC set 
up, where the occurring force has to be compensated by the 
experimental arrangement.

In both SAB and AC, the phase difference that occurs in 
interferometry is directly proportional to that of the 
ideal cases only if the initial spin state vector is an 
eigenstate of $\hat{\sigma}_{z}$. For arbitrary direction 
of the dipole, the phase differences are nonlinearly 
related to the ideal phase shifts as  $\phi = \arctan 
(\cos \theta \tan \phi_{{\text{SAB/AC}}})$ due to the 
appearance of a geometric phase associated with the 
nontrivial spin path. 

We end by noting that the precession of the dipole in SAB 
and AC is a local effect that could be obtained classically. 
In particular, it does not require a closed loop and can 
therefore be tested in a polarization measurement. Furthermore, 
for a closed space-time loop as obtained in interferometry, 
the precession effects 
can be inferred from the interference pattern characterized 
by the nonideal phase differences and the visibilities,
thereby confirming the local torque.
Extending interferometry experiments to spin-polarized 
particles in nonideal configurations would therefore be 
pertinent as they would verify SAB and AC as essentially 
local and nontopological effects. 

\section{Acknowledgments}
This work was financed by Deutsche Forschungsgemeinschaft 
(P.H.) and by the Swedish Research Council (E.S.).


\begin{thebibliography}{99}
\bibitem{aharonov59} Y. Aharonov and D. Bohm, 
{\it Phys. Rev.} {\bf 115}, 485 (1959).
\bibitem{zeilinger86} A. Zeilinger, 
``Generalized Aharonov-Bohm Experiments with Neutrons'', 
in {\it Fundamental Aspects of Quantum Theory}, 
eds. V.Gorini and A. Frigerio, NATO ASI Series B, Vol. 144, 
Plenum Press (1986), p. 311.
\bibitem{aharonov84} Y. Aharonov and A. Casher, 
{\it Phys. Rev. Lett.} {\bf 53}, 319 (1984).
\bibitem{allman92} B.E. Allman, A.Cimmino, A.G. Klein, 
G.I. Opat, H. Kaiser, and S.A. Werner, 
{\it Phys. Rev. Lett.} {\bf 68}, 2409 (1992).
\bibitem{allman93} B.E. Allman, A.Cimmino, A.G. Klein, 
G.I. Opat, H. Kaiser, and S.A. Werner, 
{\it Phys. Rev. A} {\bf 48}, 1799 (1993).
\bibitem{allman99} B.E. Allman, W.-T. Lee, O.I. Motrunich, 
and S.A. Werner,
{\it Phys. Rev. A} {\bf 60}, 4272 (1999).
\bibitem{cimmino89} A. Cimmino, G.I. Opat, A.G. Klein, 
H. Kaiser, S.A. Werner, M.Arif, and R. Clothier, 
{\it Phys. Rev. Lett.} {\bf 63}, 380 (1989).
\bibitem{badurek93} G. Badurek, H. Weinfurter, R. G\"ahler,
A. Kollmar, S. Wehinger, and A. Zeilinger, 
{\it Phys. Rev. Lett.} {\bf 71}, 307 (1993).
\bibitem{sangster93} K. Sangster, E.A. Hinds, S.M. Barnett, 
E. Riis, and A.G. Sinclair,
{\it Phys. Rev. Lett.} {\bf 71}, 3641 (1993). 
\bibitem{sangster95} K. Sangster, E.A. Hinds, S.M. Barnett, 
E. Riis, and A.G. Sinclair,
{\it Phys. Rev. A} {\bf 51}, 1776 (1995).
\bibitem{gorlitz95} A. G\"orlitz, B. Schuh, and A. Weis, 
{\it Phys. Rev. A} {\bf 51}, R4305 (1995).
\bibitem{dowling99} J.P. Dowling, C.P. Williams, and J.D. Franson, 
{\it Phys. Rev. Lett.} {\bf 83}, 2486 (1999).
\bibitem{he93} X.-G. He and B.H.J. McKellar, 
{\it Phys. Rev. A} {\bf 47}, 3424 (1993). 
\bibitem{wilkens94} M. Wilkens, 
{\it Phys. Rev. Lett.} {\bf 72}, 5 (1994).
\bibitem{peshkin99} M. Peshkin, 
{\it Found. Phys.} {\bf 29}, 481 (1999).
\bibitem{peshkin95} M. Peshkin and H.J. Lipkin,
{\it Phys. Rev. Lett.} {\bf 74}, 2847 (1995).
\bibitem{goldhaber89} A.S. Goldhaber, 
{\it Phys. Rev. Lett.} {\bf 62}, 482 (1989).
\bibitem{han92} Y.D. Han and I.G. Koh, 
{\it Phys. Lett. A} {\bf 167}, 341 (1992).
\bibitem{sjoqvist00} E. Sj\"oqvist, 
{\it Phys. Lett. A} {\bf 270}, 10 (2000).
\bibitem{jackson75} J.D. Jackson, 
{\it Classical Electrodynamics}, 
2nd ed. (Wiley, New York, 1975). 
\bibitem{goldstein80} H. Goldstein, {\em Classical Mechanics}, 
2nd ed. (Addison-Wesley, Reading, Massachusetts, 1980) p. 232.
\bibitem{pauli41} W. Pauli, 
{\it Rev. Mod. Phys.} {\bf 13}, 203 (1941). 
\bibitem{foldy52} L.L. Foldy, 
{\it Phys. Rev.} {\bf 87}, 688 (1952).
\bibitem{bjorken64} J.D. Bjorken and S.D. Drell,
{\it Relativistic Quantum Mechanics} 
(McGraw-Hill, New York, 1964) p. 8.
\bibitem{anandan89} J. Anandan
{\it Phys. Lett. A} {\bf 138}, 347 (1989).
\bibitem{he91} X.-G. He and B.H.J. McKellar,
{\it Phys. Lett. B} {\bf 256}, 250 (1991).
\bibitem{hagen90} C.R. Hagen, 
{\it Phys. Rev. Lett.} {\bf 64}, 2347 (1990).
\bibitem{pancharatnam56} S. Pancharatnam, 
{\it Proc. Indian Acad. Sci. A} {\bf 44}, 247 (1956).
\bibitem{wagh95} A.G. Wagh and V.C. Rakhecha, 
{\it Phys. Lett. A} {\bf 197}, 107 (1995). 
\bibitem{wagh98} A.G. Wagh, V.C. Rakhecha, P. Fischer, and A. Ioffe, 
{\it Phys. Rev. Lett.} {\bf 81}, 1992 (1998).
\bibitem{aharonov00} Y. Aharonov and B. Reznik,
{\it Phys. Rev. Lett.} {\bf 84}, 4790 (2000).
\bibitem{hyllus02} P. Hyllus and E. Sj\"{o}qvist,
{\it Phys. Rev. Lett.} (to appear).
\bibitem{aharonov02}
Y. Aharonov and B. Reznik, 
{\it Phys. Rev. Lett.} (to appear). 
\end{thebibliography}
\end{document}